\documentclass[english,aps,reprint]{revtex4}
\usepackage[T1]{fontenc}
\usepackage{color}
\usepackage{float}
\usepackage{amsmath}
\usepackage{graphicx}
\usepackage{amssymb}
\usepackage{esint}

\newcommand{\be}{\begin{equation}}
\newcommand{\ee}{\end{equation}}
\newcommand{\bea}{\begin{eqnarray}}
\newcommand{\eea}{\end{eqnarray}}

\newcommand{\ba}{\begin{eqnarray}}
\newcommand{\ea}{\end{eqnarray}}

\makeatletter

\@ifundefined{textcolor}{}
{%
 \definecolor{BLACK}{gray}{0}
 \definecolor{WHITE}{gray}{1}
 \definecolor{RED}{rgb}{1,0,0}
 \definecolor{GREEN}{rgb}{0,1,0}
 \definecolor{BLUE}{rgb}{0,0,1}
 \definecolor{CYAN}{cmyk}{1,0,0,0}
 \definecolor{MAGENTA}{cmyk}{0,1,0,0}
 \definecolor{YELLOW}{cmyk}{0,0,1,0}
 }

\makeatother

\makeatother

\usepackage{babel}

\makeatother

\usepackage{babel}

\makeatother

\usepackage{babel}

\begin{document}

\title{Anomalous Hall effect in superconductors with spin-orbit interaction
 }

\author{P.D. Sacramento$^1$, M.A.N. Ara\'ujo$^{1,2}$, V.R. Vieira$^1$, V.K.
Dugaev$^{1,3,4}$ and J. Barna\'s$^5$}

\affiliation{$^1$ CFIF, Instituto Superior T\'ecnico, TU Lisbon,
 Av. Rovisco Pais, 1049-001 Lisboa, Portugal}

\affiliation{$^2$ Departamento de F\'{\i}sica, Universidade de \'Evora, P-7000-671
\'Evora, Portugal
}

\affiliation{$^3$
Department of Physics, Rzesz\'ow University of Technology,
al.~Powsta\'nc\'ow Warszawy 6, 35-959 Rzesz\'ow, Poland
}

\affiliation{$^4$Institut f\"ur Physik, Martin-Luther-Universit\"at Halle-Wittenberg,
Heinrich-Damerow-St. 4, 06120 Halle, Germany}

\affiliation{
$^5$Department of Physics, Adam Mickiewicz University,
Umultowska~85, 61-614 Pozna\'n, Poland}

\begin{abstract}
%\textbf{preprint \today{}}\\
We calculate the anomalous Hall conductance of  superconductors
with spin-orbit interaction and with either uniform or local
magnetization. In the first case we consider a uniform
ferromagnetic ordering in a spin triplet superconductor, while in
the  second case we consider a conventional s-wave spin singlet
superconductor with a magnetic impurity (or a diluted set of
magnetic impurities). In the latter case we show that the
anomalous Hall conductance can be used to track the quantum phase
transition, that occurs when the spin coupling between the
impurity and electronic spin density exceeds a certain critical
value. In both cases we find that for large spin-orbit coupling
the superconductivity is destroyed and the Hall conductance
oscillates strongly.
\end{abstract}

\maketitle
\global\long\def\ket#1{\left| #1\right\rangle }

\global\long\def\bra#1{\left\langle #1 \right|}

\global\long\def\kket#1{\left\Vert #1\right\rangle }

\global\long\def\bbra#1{\left\langle #1\right\Vert }

\global\long\def\braket#1#2{\left\langle #1\right. \left| #2 \right\rangle }

\global\long\def\bbrakket#1#2{\left\langle #1\right. \left\Vert #2\right\rangle }

\global\long\def\av#1{\left\langle #1 \right\rangle }

\global\long\def\tr{\text{Tr}}

\global\long\def\im{\text{Im}}

\global\long\def\re{\text{Re}}

\global\long\def\sign{\text{sign}}

\global\long\def\Det{\text{Det}}

%\tableofcontents{}\newpage{}

\section{Introduction}

Anomalous Hall effect (AHE) was observed in metallic ferromagnets
long time ago as a Hall current generated by electric field in the
absence of external magnetic field \cite{hall1880,chien79}. Since
then several physical mechanisms of the AHE have been proposed,
related to the side-jump and skew scattering from impurities
\cite{smit58,berger72,crepieux01_1}, inhomogeneous internal
magnetization \cite{taguchi01,bruno04}, internal spin-orbit
interaction \cite{karplus54}, and topology of electron energy
bands \cite{jungwirth02,m_onoda02}. Theory of AHE attracted much
attention recently
\cite{crepieux01_1,sinova04,haldane04,nagaosa10} because it
reveals some very unusual properties of solids such as the
existence of monopoles in the momentum space or generation of
topological gauge fields.

One of the most intriguing models of AHE is the one based on an
intrinsic mechanism \cite{jungwirth02,m_onoda02} related to the
nontrivial topology of electron energy bands. In frame of this
mechanism, the main contribution to AHE is due to electron states
well below the Fermi energy \cite{streda82}. The simplest model,
in which this mechanism of AHE can be realized, is the model of a
magnetized two-dimensional electron gas with Rashba spin-orbit
(SO) interaction \cite{rashba84}. Unfortunately, it turns out that
if the system is in the metallic state, i.e., if there is no gap
at the Fermi surface, then the contribution of electron states at
the Fermi surface can totally compensate the other contributions
so that the resulting off-diagonal conductivity is zero
\cite{nunner07}. In the opposite case, when the chemical potential
lies in the gap, the anomalous Hall conductivity $\sigma _{xy}$ is
nonzero and quantized in units of $e^2/h$. The theory of quantized
AHE is quite similar to the theory of integer quantum Hall effect,
where the gap is due to the Landau quantization in a strong
magnetic field \cite{thouless82,niu85}.

In this work we consider a 2D electron gas with nonzero
magnetization and Rashba SO interaction. Such a model was used
earlier for a description of intrinsic AHE \cite{nunner07}.
However, we calculate the AHE in the case when the electron system
is additionally superconducting. The superconductivy produces a
gap at the Fermi level, suppressing the contribution to AHE from
the Fermi surface. Thus, one can expect that only the filled
electronic states below the gap contribute to the AHE. The
possibility of AHE in superconductors has been already considered
in the case of ferromagnet-superconductor double tunnel junctions
\cite{takahashi02}, where side jump and/or skew scattering from
impurities have been assumed as possible physical mechanisms
responsible for the effect. This is, however,  essentially
different from our model, where we consider the intrinsic
mechanism of AHE. Since in a superconductor charge is not
conserved due to the particle-hole mixture, we do not expect any
quantization of the anomalous Hall conductance. This was already
shown for the usual Hall conductance in conventional
superconductors in very high magnetic fields, where the Landau
level description is appropriate \cite{Hall}.

Various materials are known to show the coexistence of
ferromagnetism and superconductivity
\cite{Fertig,Ishikawa,Felner,Bernhard,Saxena,Aoki,Huy,Guanghan}
and, in particular, the presence of spin-orbit interaction due to
the lack of spatial inversion symmetry
\cite{Bauer,Samokhine,Akazawa,Kimura}. We note, that the
possibility of magnetoelectric effects in non-centrosymmetric
superconductors was predicted already long time ago
\cite{Edelstein1}, where it was shown that a supercurrent should
induce a spin polarization and  reversely a Zeeman-like term
should induce a supercurrent \cite{Yip1} as a result of strong
spin-orbit interaction. Other effects due to the interplay of
ferromagnetism and superconductivity have also been considered
\cite{Edelstein2,Samokhin,Fujimoto1}. Recently, the interplay
between superconductivity, magnetism and spin-orbit interaction
(or topological insulators \cite{Hasan,Qi}) has  received
additional attention due to the possibility of Majorana edge
states in a finite system or inside superconducting vortices
\cite{Fu,Sau,Sato}, with its possible applications in topological
quantum computation. Moreover, the coexistence of magnetism and
superconductivity turned out to be interesting also from the point
of view of possible applications in spintronics
\cite{takahashi,yokoyama}.

 In this work we consider in section II a spin
triplet superconductor while  in section III a conventional
superconductor with a magnetic impurity \cite{Sakurai}. In both
cases we analyze the influence of Rashba spin-orbit interaction.
In the first case the magnetization is due to a ferromagnetic
order, whereas in the second case the system is locally polarized
by a magnetic impurity. The latter situation may also be achieved
when considering a superconducting film with a magnetic dot
justaposed. It has been shown before that if the coupling between
the magnetic impurity and the spin density of conduction electrons
is strong enough, the system becomes magnetized through a first
order quantum phase transition \cite{Morr,us} that leads to
discontinuities in various physical quantities \cite{QI}. In both
cases we calculate the anomalous Hall conductance. We show that
the Hall conductance of a superconductor with the magnetic
impurity can be used to reveal the quantum phase transition.
Finally, we conclude with section IV.

\section{AHE in a triplet superconductor}

We consider first a superconductor with a uniform magnetization.
Since magnetism and superconductivity compete, a spin singlet
superconductor is not stable due to Cooper pair breaking.
Therefore, we consider a spin triplet superconductor, where
magnetism and superconductivity can coexist. The system  is
described by the tight-binding model in two dimensions, to which
we add a superconducting pairing term with the appropriate
symmetry. Additionally, we also include the Rashba spin-orbit term
\cite{rashba84}, which is generally allowed in non-centrosymmetric
materials. Due to the spin-orbit term, a spin singlet component,
$\Delta_s$, is generally induced and therefore there is a pairing
mixture in the system \cite{Rashba}.

We write the electron operators, $\psi_{\vec{k},\sigma}$, in terms
of the Bogoliubov operators, $\gamma_{n,\vec{k}}$, as \be
\psi_{\vec{k},\sigma} = \sum_n \left( u_n(\vec{k},\sigma)
\gamma_{n,\vec{k}} - \sigma v_n(\vec{k},\sigma)^*
\gamma_{n,-\vec{k}}^{\dagger} \right), \ee where $\vec{k},n$ label
the eigenstates of the system. The wave functions and energy
eigenvalues satisfy the Bogoliubov -- de Gennes equations
\cite{deGennes}, which can be written as
\begin{equation}
\label{bdgtriplet} \left(\begin{array}{cccc}
\epsilon_{\vec{k}}-h_z & \alpha (\sin k_y+i \sin k_x) & -d_x+i d_y & d_z+\Delta_s \\
\alpha (\sin k_y-i \sin k_x) & \epsilon_{\vec{k}}+h_z & d_z-\Delta_s & d_x+i d_y  \\
-d_x -i d_y & d_z-\Delta_s & -\epsilon_{\vec{k}}+h_z & \alpha (\sin k_y-i \sin k_x)  \\
d_z+\Delta_s & d_x -i d_y & \alpha (\sin k_y+i \sin k_x) & -\epsilon_{\vec{k}}-h_z
\end{array}\right)
\left(\begin{array}{c}
u_n(\vec{k},\uparrow) \\
u_n(\vec{k},\downarrow) \\
v_n(-\vec{k},\uparrow) \\
v_n(-\vec{k},\downarrow) \\
\end{array}\right)= \epsilon_{\vec{k},n}
\left(\begin{array}{c}
u_n(\vec{k},\uparrow) \\
u_n(\vec{k},\downarrow) \\
v_n(-\vec{k},\uparrow) \\
v_n(-\vec{k},\downarrow) \\
\end{array}\right).
\end{equation}
Here, $\epsilon_{\vec{k}}=-2 t (\cos k_x + \cos k_y )-\epsilon_F$
is the kinetic part, where $t$ denotes the hopping parameter set in
the following as the energy scale, $t=1$, $\epsilon_F$ is the
chemical potential, chosen in the following as $\epsilon_F=-1$,
$\vec{k}$ is a wave vector in the $xy$ plane, and we have taken
the lattice constant to be unity, $a=1$. Furthermore, $h_z$ in
Eq.(2) is the magnetization, in energy units, along the $z$ direction, while the
vector $\vec{d}=(d_x,d_y,d_z)$ is the vector representation of the
superconducting pairing ($p$-wave). Finally, the Rashba spin-orbit
term is written as $H_R = \vec{s} \cdot \vec{\sigma} = \alpha
\left( \sin k_y \sigma_x - \sin k_x \sigma_y \right)$, where
$\alpha$ is measured in the energy units, and $\sigma_x,\sigma_y$
are the Pauli matrices.

\begin{figure}[t]
\begin{centering}
\includegraphics[width=0.35\columnwidth]{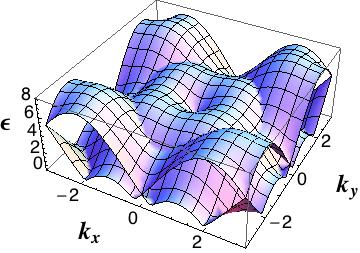}
\includegraphics[width=0.35\columnwidth]{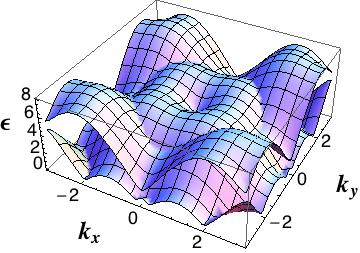}
\includegraphics[width=0.35\columnwidth]{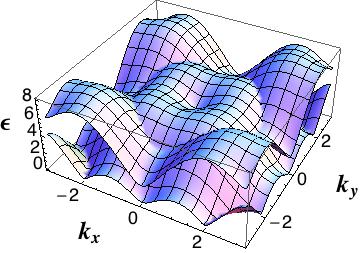}
\includegraphics[width=0.35\columnwidth]{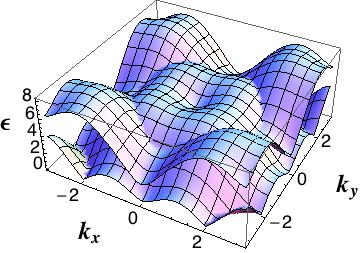}
\par\end{centering}
\caption{\label{fig1} Energy bands in units of the hopping, $t$, as a function of momenta
$k_x,k_y$ in the normal phase for
$\lambda_{so}=2$ and various values of the magnetization: from
left to right $h_z=0$ and $h_z=0.5$ (top); $h_z=1$ and $h_z=1.2$
(bottom). }
\end{figure}

The pairing matrix can be written as \cite{Sigrist}
\begin{equation}
\label{bdgtriplet2} \Delta=
\left(\begin{array}{cc}
\Delta_{\uparrow,\uparrow} & \Delta_{\uparrow,\downarrow} \\
\Delta_{\downarrow,\uparrow} & \Delta_{\downarrow,\downarrow}
\end{array}\right)
= \left(\begin{array}{cc}
-d_x+i d_y & d_z \\
d_z & d_x +i d_y
\end{array}\right).
\ee Thus, we can write $d_x=\left(
\Delta_{\downarrow,\downarrow}-\Delta_{\uparrow,\uparrow}
\right)/2$, $d_y=-i \left(
\Delta_{\downarrow,\downarrow}+\Delta_{\uparrow,\uparrow}
\right)/2$, and $d_z=\Delta_{\uparrow,\downarrow}$, while  the
vector $\vec{q}=i \vec{d} \times \vec{d}^*$ is given by $q_x = Re
\left[ \left( \Delta_{\downarrow,\downarrow} +
\Delta_{\uparrow,\uparrow} \right) \Delta_{\uparrow,\downarrow}^*
\right]$, $q_y = Im \left[ \left( \Delta_{\downarrow,\downarrow} -
\Delta_{\uparrow,\uparrow} \right) \Delta_{\uparrow,\downarrow}^*
\right]$ and $q_z = \frac{1}{2} \left[
|\Delta_{\uparrow,\uparrow}|^2 -
 |\Delta_{\downarrow,\downarrow}|^2 \right]$.
When this vector vanishes, the pairing is called unitary. We have
verified that considering the s-wave component has  generally a
very small effect on our results,  and therefore we assume
$\Delta_s=0$ in the following.

The energy eigenvalues of Eq. \ref{bdgtriplet} can be written (for
$\Delta_s=0$) as \be \label{bands} \epsilon_{\vec{k},\alpha_1,\alpha_2}
= \alpha_1 \sqrt{z_1 +\alpha_2 2 \sqrt{z_2}}, \ee  where \bea
z_1 &=& \vec{d}\cdot \vec{d} + \vec{s}\cdot \vec{s} + \epsilon_{\vec{k}}^2 + h_z^2 \nonumber \\
z_2 &=& \left( \vec{d}\cdot \vec{s} \right)^2 +
(\epsilon_{\vec{k}}^2+d_z^2)(\vec{s}\cdot \vec{s} + h_z^2 ), \eea and
$\alpha_1,\alpha_2=\pm$.

In the normal phase ($\vec{d}=0$), the spin-orbit coupling lifts
the spin degeneracy of the energy bands in the tight-binding
model, except at $\vec{k}=(0,0)$, $(\pi,\pi)$ and $(0,\pi)$ (and
equivalent points). These remaining degeneracies are lifted when
including the magnetization. This is shown in Fig. \ref{fig1},
where the two energy bands are shown as a function of momentum for
$\lambda_{so}=\alpha /2=2$ and various values of $h_z$. As can be
seen from Eq. \ref{bands}, the lowest band is gapless at the
points where \be \left( \vec{s}\cdot \vec{s} + h_z^2 \right) +
\epsilon_{\vec{k}}^2 = 2 \sqrt{ \left( \vec{s}\cdot \vec{s} +
h_z^2 \right) \epsilon_{\vec{k}}^2 }. \ee In a general case
($\vec{d}\ne 0$), the lowest band has gapless points that are
solutions of the equation $z_1=2 \sqrt{z_2}$, which yields \be
\vec{d}\cdot \vec{d} + \vec{s}\cdot \vec{s} + \epsilon_{\vec{k}}^2
+ h_z^2 = 2 \sqrt{ \left( \vec{d}\cdot \vec{s} \right)^2 +
(\epsilon_{\vec{k}}^2+d_z^2)(\vec{s}\cdot \vec{s} + h_z^2 ) }. \ee
Thus, in the superconducting phase the system is generally gapped.
In particular, without the spin-orbit interaction the gapless
points are obtained by $\vec{d} \cdot \vec{d} +
\epsilon_{\vec{k}}^2 =0$ which implies particular values for the
chemical potential.

The charge current along a link in the lattice can be obtained by
adding a vector potential to the kinetic and spin-orbit terms and
taking a functional derivative of the Hamiltonian with respect to
the vector potential \cite{deGennes,Durst}, or through its
definition in the charge continuity equation \cite{BTK}. The
zero-momentum charge current in the $\mu=x,y$ direction can be
written as \be j_{\mu} = \sum_{\vec{k}}
\bar{\psi}_{\vec{k}}^{\dagger} V_{\vec{k}}^{\mu}
\bar{\psi}_{\vec{k}}, \ee where $\bar{\psi}_{\vec{k}} =
\left(\begin{array}{cc} \psi_{\vec{k},\uparrow} &
\psi_{\vec{k},\downarrow}
\end{array}\right)^T$,
and
\bea
V^x &=& \frac{2 e}{\hbar} \left( -t \eta_{\vec{k},-}^x I + \lambda_{so} \eta_{\vec{k},+}^x \sigma_y
\right) \nonumber \\
V^y &=& \frac{2 e}{\hbar} \left( -t \eta_{\vec{k},-}^y I -
\lambda_{so} \eta_{\vec{k},+}^y \sigma_x \right) \eea is a
velocity matrix operator \cite{Vafek}. Here
$\eta_{\vec{k},+}^{\mu} = \cos \left( \vec{k} \cdot
\vec{\delta}_{\mu} \right)$ and $\eta_{\vec{k},-}^{\mu} = \sin
\left( \vec{k} \cdot \vec{\delta}_{\mu} \right) $, where
$\vec{\delta}_{\mu}$ is a vector displacement (in units of the
lattice constant) between nearest-neighbors along the $\mu$
direction. In turn, $I$ is the $2\times 2$ unit matrix.

\begin{figure}[t]
\begin{centering}
\includegraphics[width=0.35\columnwidth]{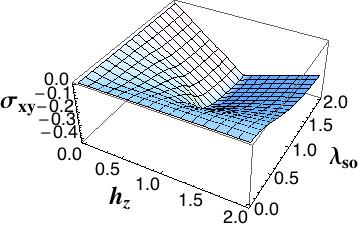}
\par\end{centering}
\caption{\label{fig2} Anomalous Hall conductance in units of $e^2/h$ in the normal
phase as a function of $h_z$ and $\lambda_{so}$. }
\end{figure}

The Hall conductance can be now calculated using a Kubo like
formula \cite{Mahan}, which in the limit of uniform and stationary
current, $\vec{q} \rightarrow 0$ and $ \omega \rightarrow 0$, is
given by \bea Re (\sigma_{xy}) &=& -i \frac{\hbar}{N}
\sum_{\vec{k}} \sum_{\alpha,\beta} \sum_{\gamma,\delta} \sum_{n,m}
\frac{f_{n,\vec{k}}-f_{m,\vec{k}}}{(\epsilon_{n,\vec{k}}-
\epsilon_{m,\vec{k}}+i 0^+)^2}    \nonumber \\
& & \left( V_{\vec{k};\alpha,\beta}^x V_{\vec{k};\gamma,\delta}^y
u_n(\vec{k},\alpha)^* u_n(\vec{k},\delta) u_m(\vec{k},\beta) u_m(\vec{k},\gamma)^* \right.  \nonumber \\
& - &  \left. V_{\vec{k};\alpha,\beta}^x
V_{-\vec{k};\gamma,\delta}^y \gamma \delta u_n(\vec{k},\alpha)^*
v_n(-\vec{k},\gamma) u_m(\vec{k},\beta) v_m(-\vec{k},\delta)^*
\right). \eea where $N$ is the number of sites and $f_{n,\vec{k}}$
is the Fermi function for the state described by $n$ and
$\vec{k}$. In the normal phase the wave functions $u$ and $v$ are
decoupled. The presence of superconducting pairing mixes the
particle and hole character and, as already mentioned above,
charge is no longer a good quantum number. The results for the
Hall conductance depend then on the choice of the pairing matrix
\cite{Sigrist,Volovik}.

Let us now assume that the pairing amplitude is a free parameter.
This describes the situations where superconductivity is induced
by proximity and therefore no self-consistent solution is implied.
This also applies to a situation where $\sigma_{xy}$ is measured on a
normal sample in which superconductivity pairing exists due to proximity
effect in the presence of a nearby triplet superconductor. 
We consider both unitary and nonunitary cases. Then we consider
the case, where the pairing amplitude is determined by solving the
Bogoliubov -- de Gennes equations self-consistently. In the latter
case we consider a nonunitary situation, for which the amplitudes
$\Delta_{\uparrow,\uparrow}$ and $\Delta_{\downarrow,\downarrow}$
are real, to simplify. This in turn implies that $d_y$ is
imaginary. In all cases we take $\Delta_{\uparrow,\downarrow}=0$
($d_z=0$), which means that only the $q_z$ component may be
nonvanishing.

\begin{figure}[t]
\begin{centering}
\includegraphics[width=0.35\columnwidth]{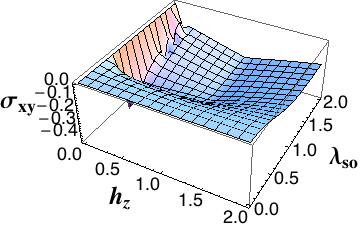}
\includegraphics[width=0.35\columnwidth]{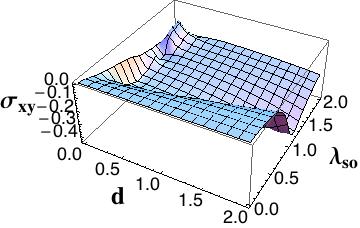}
\includegraphics[width=0.35\columnwidth]{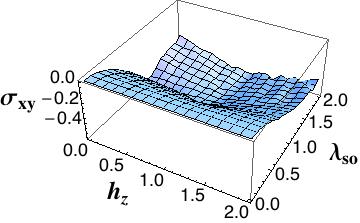}
\includegraphics[width=0.35\columnwidth]{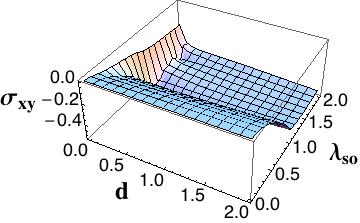}
\par\end{centering}
\caption{\label{fig3} Anomalous Hall conductance for a spin
triplet superconductor. Left panels present Hall conductance  as a
function of $h_z$ and $\lambda_{so}$ for $d=1$, whereas right
panels as a function of $d$ and $\lambda_{so}$ for $h_z=0.5$. Top
figures correspond to the unitary case, see Eq. \ref{eq1}, while
bottom figures correspond to the nonunitary case, Eq. \ref{eq2}. }
\end{figure}

In Fig. \ref{fig2} the anomalous Hall conductance in the normal
phase (zero pairing amplitude) is plotted as a function of the
magnetization $h_z$ and spin-orbit coupling $\lambda_{so}$. The
Hall conductance vanishes if either the magnetization or the
spin-orbit coupling vanishes. Then, the absolute value of the Hall
conductance increases as either parameter increases. Dependence on
$h_z$ is more complex, as the conductance reaches a minimum around
$h_z=1\sim -\epsilon_F$, which shifts if we change the chemical
potential. The minimum in the Hall conductance as a function of
the magnetization $h_z$ (keeping the spin-orbit constant, for
instance $\lambda_{so}=2$) is associated with the gaplessness of
the spectrum at the point $(0,\pi)$ and the equivalent points (see
also Fig. \ref{fig1}).

Now, we consider the superconducting phase. Since the spin-orbit
coupling renders the type of pairing undefined (with the mixture
of spin triplet and spin singlet pairings), the strength of the
triplet pairing is expected to be weakened in comparison to the
same superconductor with a vanishing spin-orbit coupling. However,
it has been shown before \cite{Frigeri} that the amplitude of the
triplet pairing is not affected by the spin-orbit term when the
vector $\vec{d}$ is parallel to the spin-orbit vector $\vec{s}$.
We have found that this pairing choice leads to results for the
anomalous Hall conductance, that are very similar to those for the
Hall conductance in the normal phase. This indicates that for this
particular case, the superconducting order does not change
significantly the Hall conductance, and therefore we do not show
the corresponding results. We have also considered other choices
of pairing, for which the vector $\vec{d}$ is not parallel to the
spin-orbit vector $\vec{s}$. We have considered both unitary and
non-unitary cases. It is already known for a  unitary case
\cite{Frigeri}, that even though the amplitude of the triplet
coupling is somewhat weakened with respect to the case of
vanishing spin-orbit term, it is still finite.

In Fig. \ref{fig3} we show the anomalous Hall conductance in the
superconducting phase for the two choices of the triplet pairing.
We consider a unitary choice given by \bea
\Delta_{\uparrow,\uparrow}&=&d(-\sin k_y + i \sin k_x),
\Delta_{\uparrow,\downarrow}= \Delta_{\downarrow,\uparrow}=0,
\Delta_{\downarrow,\downarrow}=d(\sin k_y + i \sin k_x) \nonumber \\
q_x &=&0, q_y=0, q_z=0.
\label{eq1}
\eea
and a nonunitary choice given by
\bea
\Delta_{\uparrow,\uparrow}&=&d \sin k_x,
\Delta_{\uparrow,\downarrow}= \Delta_{\downarrow,\uparrow}=0,
\Delta_{\downarrow,\downarrow}=0 \nonumber \\
q_x &=&0, q_y=0, q_z=\frac{d^2}{2} \sin^2 k_x. \label{eq2} \eea In
the case of unitary coupling (top panels of Fig. \ref{fig3}),
$\sigma_{xy}=0$ if either $\lambda_{so}=0$ or $h_z=0$. However, in
the case of a non-unitary coupling (bottom panels of Fig.
\ref{fig3}), $\sigma_{xy}=0$ if $\lambda_{so}=0$, but for a
nonzero spin-orbit coupling there is a finite Hall conductance
even if $h_z=0$. In this nonunitary case there is a magnetization
induced by the pairing, which leads to a finite $\sigma_{xy}$ in a
similar way as in $^3He$.

In the unitary case the energy spectrum has a gap at the Fermi
energy. This gap decreases as $\lambda_{so}$ increases. As
$\lambda_{so}$ grows, the gap between the first and the second
bands seems to decrease slightly and then it increases. In
general, one can expects that small gaps between the bands will
lead to large contributions to the Hall conductance. In the
nonunitary case the energy spectrum also has a gap at the Fermi
surface, which is small for small $\lambda_{s0}$, increases for
slightly larger spin-orbit coupling, but vanishes when
$\lambda_{so}$ exceed $\lambda_{so}\sim 0.7$. As $\lambda_{so}$
grows further the gap between the first and second bands
increases.

\begin{figure}[t]
\begin{centering}
\includegraphics[width=0.35\columnwidth]{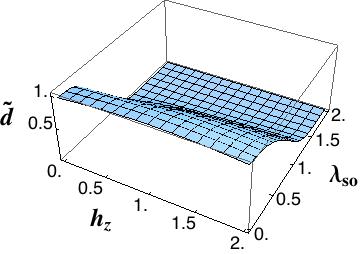}
\includegraphics[width=0.4\columnwidth]{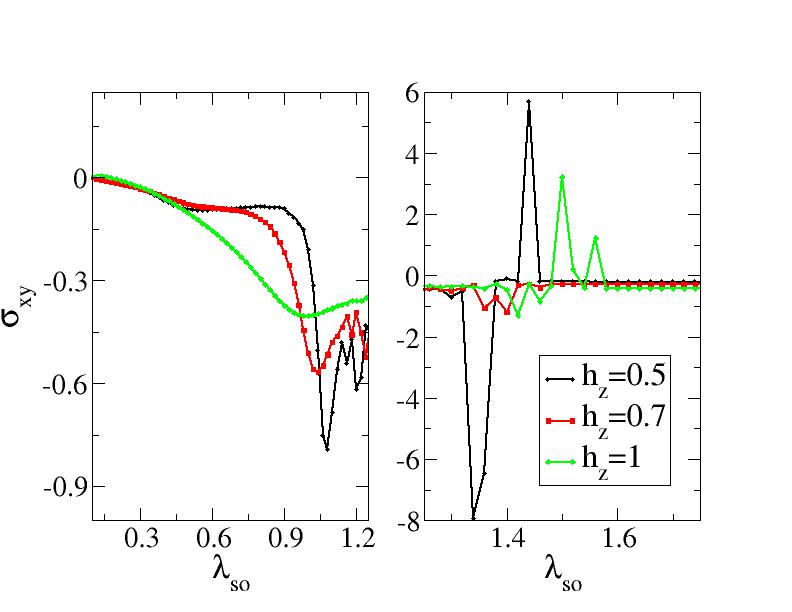}
\par\end{centering}
\caption{\label{fig4}Spin triplet superconductor calculated
self-consistently. Left: $\tilde{d}$ as a function of $h_z$ and
$\lambda_{so}$. Right: Hall conductance along cuts of constant
magnetization as a function of the spin-orbit coupling. }
\end{figure}

In the case when the superconductivity is intrinsic to the material,
we have to solve the Bogoliubov -- de Gennes equations
self-consistently. We look for a situation of the type
\bea
\Delta_{\uparrow,\uparrow}&=&\tilde{d}(-\sin k_x + \sin k_y),
\Delta_{\uparrow,\downarrow}= \Delta_{\downarrow,\uparrow}=0,
\Delta_{\downarrow,\downarrow}=\tilde{d}(\sin k_x + \sin k_y), \nonumber \\
q_x=0, q_y &=&0, q_z=\frac{\tilde{d}^2}{2} (-4 \sin k_x \sin k_y),
\label{eq3} \eea where the amplitude $\tilde{d}$ is determined
self-consistently for a given magnetization, taking into account
that \be \tilde{d}=\frac{g}{N} \sum_{\vec{k}} (-\sin k_x + \sin
k_y) \langle \psi_{\vec{k} \uparrow} \psi_{-\vec{k} \uparrow}
\rangle . \ee where $g$ is the pairing interaction. The
corresponding numerical results are shown in Fig. \ref{fig4}. As
the left panel shows, the superconductivity is destroyed for large
enough spin-orbit coupling. In the right panel we see that the
Hall conductance (as a function of $\lambda_{so}$) decreases with
increasing $\lambda_{so}$, and as the transition to the normal
phase appears, there are oscillations  of the Hall conductance
with relatively large amplitudes.

\section{AHE in a conventional superconductor with magnetic impurity}

Consider now a classical spin immersed in a two-dimensional
$s$-wave conventional superconductor. We use now a description of
the system in the real space. In the center of the system,
$\vec{r}=\vec{l}_c=(x_c,y_c)$, we place a classical spin along the
$z$ direction. The kinetic energy part is  described by a
tight-binding model with hopping amplitude $t$, similarly as in
the case of triplet superconductivity. The superconductor pairing
is taken as $s$-wave, and the spin-orbit interaction \cite{Pareek}
is assumed as in the preceding section. The electron operator is
written in terms of the Bogoliubov operators, \be
\psi(\vec{r},\sigma) = \sum_n \left( u_n(\vec{r},\sigma) \gamma_n
-\sigma v_n(\vec{r},\sigma)^* \gamma_n^{\dagger} \right). \ee

The zero momentum charge current in the $\mu=x,y$ direction can be
written as $j_{\mu} = \sum_{\vec{r}}
\bar{\psi}_{\vec{r}}^{\dagger} V^{\mu} \bar{\psi}_{\vec{r}}$,
where $\bar{\psi}_{\vec{r}} = \left(\begin{array}{cc}
\psi_{\vec{r},\uparrow} & \psi_{\vec{r},\downarrow}
\end{array}\right)^T$,
and the velocity matrix operators are given by
\bea
V^x &=& \frac{e}{\hbar} \left( i t \eta_{-}^x I + \lambda_{so} \eta_{+}^x \sigma_y
\right) \nonumber \\
V^y &=& \frac{e}{\hbar} \left( i t \eta_{-}^y I - \lambda_{so}
\eta_{+}^y \sigma_x \right). \eea Here $f(\vec{r}) \eta_{+}^{\mu}
g(\vec{r}) = f(\vec{r}+\vec{\delta}_{\mu}) g(\vec{r}) + f(\vec{r})
g(\vec{r}+\vec{\delta}_{\mu})$ and $f(\vec{r}) \eta_{-}^{\mu}
g(\vec{r}) = f(\vec{r}+\vec{\delta}_{\mu}) g(\vec{r}) - f(\vec{r})
g(\vec{r}+\vec{\delta}_{\mu}) $, where $\vec{\delta}_{\mu}$ is a
displacement between nearest-neighbors along the $\mu$ direction,
while $\sigma_x,\sigma_y$ are Pauli matrices, as above.

The real space wave functions obey the Bogoliubov -- de Gennes
equations for the energy excitations $\epsilon_n$ \be \left( \begin{array}{cccc} -h -\epsilon_F -J
\delta_{\vec{r},\vec{l}_c}  & \Delta_{\vec{r}} & \lambda_{so}
(-\eta_x + i \eta_y ) & 0
 \\
\Delta_{\vec{r}}^* & h +\epsilon_F -J \delta_{\vec{r},\vec{l}_c}   & 0 & \lambda_{so} (\eta_x - i \eta_y )
 \\
\lambda_{so} (\eta_x + i \eta_y ) & 0  &
-h -\epsilon_F +J \delta_{\vec{r},\vec{l}_c} & \Delta_{\vec{r}} \\
0 & \lambda_{so} (-\eta_x - i \eta_y ) & \Delta_{\vec{r}}^*  & h +\epsilon_F +
J \delta_{\vec{r},\vec{l}_c}
 \end{array} \right)
\left( \begin{array}{c}
u_n(\vec{r},\uparrow) \\
v_n(\vec{r},\downarrow) \\
u_n(\vec{r},\downarrow) \\
v_n(\vec{r},\uparrow) \end{array} \right)
= \epsilon_n
\left( \begin{array}{c}
u_n(\vec{r},\uparrow) \\
v_n(\vec{r},\downarrow) \\
u_n(\vec{r},\downarrow) \\
v_n(\vec{r},\uparrow) \end{array} \right), \ee
%\begin{multicols}{2}
\noindent where $h=t\hat{s}_{\vec{\delta}}$ with
$\hat{s}_{\vec{\delta}} f(\vec{r})=f(\vec{r}+\vec{\delta})$.
Furthermore, $\eta_x=\pm 1$ if the neighbor along $x$ is $i_x+1$
($i_x-1$) and $\eta_y=\pm 1$ if the neighbor along $y$ is $i_y+1$
($i_y-1$). The parameter $J$ describes coupling between the
impurity spin and the spin density of conduction electrons. Note
that the solution of this problem requires diagonalization of a
$4N\times 4N$ matrix, where $N$ is the number of lattice sites.
This is in contrast to the problem of triplet superconductor
described in the previous section, where a partial diagonalization
was possible due to the translational invariance. Owing to this
symmetry, the problem could be reduced  to a simple
diagonalization of a $4\times 4$ matrix for each momentum value.
Since the effect of the magnetic impurity is rather local, a
system of $15\times 15$ lattice sites is sufficient to have small
finite size effects, as we have shown previously \cite{us}.
We solve the problem self-consistently, as in a previous study (see Ref.
\cite{us} for details).

As in the case of triplet superconductivity studied in section II,
the Hall conductance can be obtained from a Kubo like formula,
which now reads \bea Re (\sigma_{xy}) &=& i \frac{\hbar}{V}
\sum_{\vec{r}_1,\vec{r}_2} \sum_{\alpha,\beta}
\sum_{\gamma,\delta}
\sum_{n,m} \frac{f_{n}-f_{m}}{(\epsilon_{n}-\epsilon_{m}+i 0^+)^2}    \nonumber \\
& & \left( V_{\vec{r}_1;\alpha,\beta}^x \bar{V}_{\vec{r}_2;\gamma,\delta}^y
u_n(\vec{r}_1,\alpha)^* u_n(\vec{r}_2,\delta) u_m(\vec{r}_1,\beta) u_m(\vec{r}_2,\gamma)^* \right.  \nonumber \\
& - &  \left. V_{\vec{r}_1;\alpha,\beta}^x V_{\vec{r}_2;\gamma,\delta}^y \gamma \delta
u_n(\vec{r}_1,\alpha)^* v_n(\vec{r}_2,\gamma) u_m(\vec{r}_1,\beta) v_m(\vec{r}_2,\delta)^* \right)
\eea
In this expression $\bar{V}$ is the complex conjugate, and $V_{\vec{r}_i}^{\mu}$ means
that the operator acts on the coordinate $\vec{r}_i$.

\begin{figure}[t]
\begin{centering}
\includegraphics[width=0.3\columnwidth]{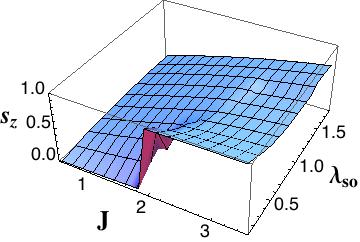}
\includegraphics[width=0.3\columnwidth]{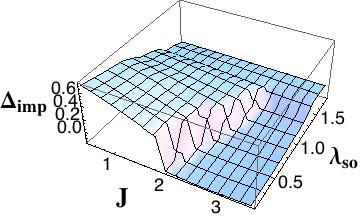}
\includegraphics[width=0.3\columnwidth]{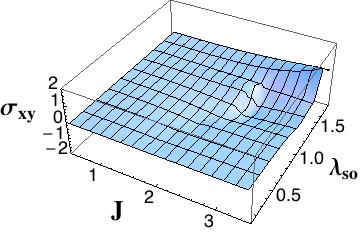}
\includegraphics[width=0.4\columnwidth]{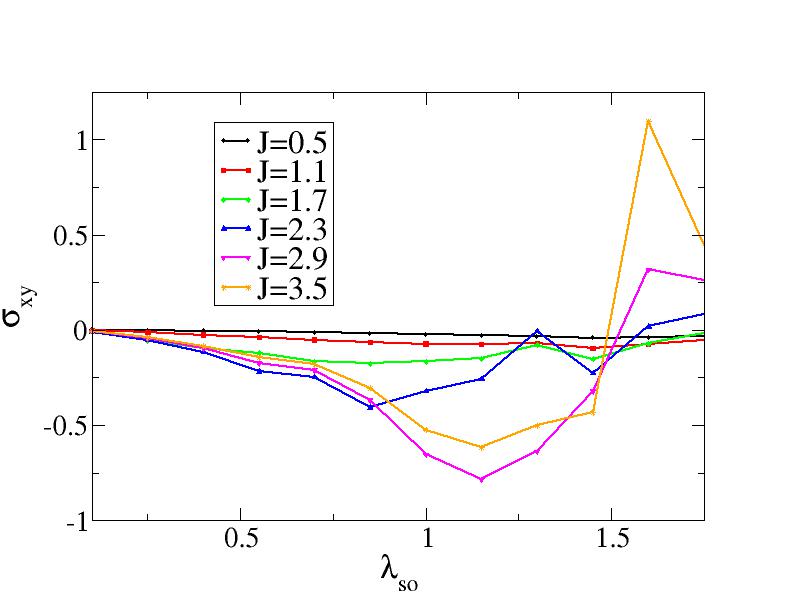}
\par\end{centering}
\caption{\label{fig5} In the top panels from left to right total magnetization,
order parameter at the impurity site, and anomalous Hall
conductance for the conventional superconductor with a magnetic
impurity, calculated for a finite system including $15\times 15$
lattice points as a function of $J$ and $\lambda_{so}$. In the lower panel
we show some cuts of the Hall conductance for $J$ fixed and
varying spin-orbit coupling.
 }
\end{figure}

\begin{figure}[t]
\begin{centering}
\includegraphics[width=0.45\columnwidth]{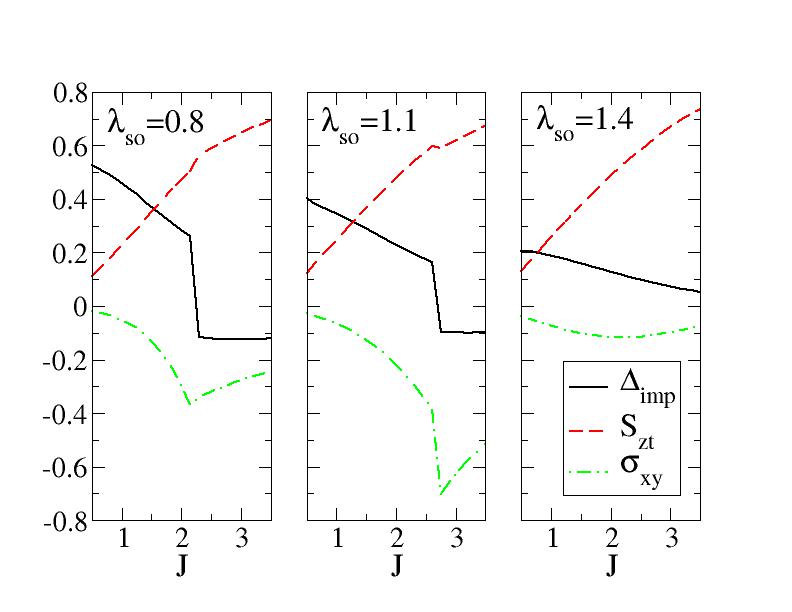}
\par\end{centering}
\caption{\label{fig6} Order parameter at the impurity site, total
magnetization, and anomalous Hall conductance for the conventional
superconductor as a function of coupling strength to the impurity
spin, calculated for the spin-orbit coupling corresponding to
$\lambda_{so}=0.8$, $\lambda_{so}=1.1$, and $\lambda_{so}=1.4$, as
indicated. The first two values cross the quantum phase transition
and for the highest value the transition turns into a crossover. }
\end{figure}

The corresponding numerical results are presented in Fig.
\ref{fig5}, where we show the total magnetization, order parameter
at the impurity site, and the anomalous Hall conductance as a
function of $J$ and $\lambda_{so}$ for a system of $15\times 15$
lattice sites. The case with no spin-orbit coupling
($\lambda_{so}=0$) was studied before \cite{Sakurai}. Note, that
the spin-orbit interaction shifts the critical value $J_c$, at
which the quantum phase transition to a magnetic state occurs, to
higher values. However, if $\lambda_{so}$ is large enough, the
transition is washed out. Note also that the various quantities
reveal the quantum phase transition when fixing $\lambda_{so}$ and
plotting them as a function of $J$. At the transition point, the
impurity "captures" one electron and breaks a Cooper pair. Note
that there is still a transition when we introduce the spin-orbit
coupling, but one needs a larger coupling parameter $J$ as the
spin-orbit increases. If we increase the spin-orbit coupling
further, superconductivity is destroyed and the Hall conductance
exhibits strong oscillations, as in the case of the triplet
superconductor.

In order to emphasize the connection between  behavior of the Hall
conductance and the quantum phase transition, we show  in Fig.
\ref{fig6} (for different spin-orbit couplings) the Hall
conductance, amplitude of the order parameter at the impurity
location, and the total magnetization of the conduction electrons
as a function of the coupling between the spin density of the
conduction electrons and the impurity spin. At the quantum phase
transition both the amplitude of the order parameter and the total
magnetization have discontinuities. At this critical coupling the
Hall conductance has a sharp minimum which therefore signals the
phase transition.

\section{Summary}

We have analyzed the anomalous Hall effect in superconductors,
considering only the intrinsic mechanism that results from the
interplay of Rashba spin-orbit interaction and magnetization. In
the normal phase, the effect appears when both spin-orbit term and
magnetization are nonzero. In a conventional spin-singlet
superconductor of $s$-wave symmetry, an extended magnetization
destroys the superconductivity. As we have shown, to have a
nonvanishing anomalous Hall conductance in the superconducting
phase it is then sufficient to assume a single magnetic impurity
in the presence of spin-orbit interaction. However, vanishing
coupling between the conduction electrons and magnetic impurity,
or vanishing spin-orbit coupling, lead to zero Hall conductance.

The case of a spin triplet superconductor is qualitatively
different. An extended magnetization does not destroy the
superconducting order. The magnetization, generally,  either can
be induced by an adjacent  ferromagnet owing to the  proximity
effect (we may also consider the superconducting order as a
proximity effect in heterostructures where some metal is coupled
to a magnet and a superconductor), or it may be an intrinsic
property of the material (described by a self-consistent solution
for the pairing amplitude). In the first case, two pairing forms
lead to different results. If the pairing is unitary, the results
are similar to those for the normal phase, and both magnetization
and spin-orbit coupling are required for a finite Hall
conductance. The superconducting case is very similar to the
normal phase also when  $\vec{d}$ is parallel to $\vec{s}$. In the
nonunitary case, however, there is a polarization associated with
the pairing amplitude, and the Hall conductance is finite as long
as the spin-orbit coupling is finite (a nonunitary pairing leads
to a finite magnetization, as in the case of $^3He$).

Since the spin-orbit interaction generates spin flips, its
moderate values destroy superconductivity in both the conventional
and triplet superconductors. In the case of $s$-wave
superconductors, critical values of the spin-coupling $J$ in the
presence of spin-orbit coupling are larger than those for zero
spin-orbit coupling, shifting thus the point at which the quantum
phase transition appears. In the case of spin triplet
superconductors with the pairing amplitude determined
self-consistently, the spin-orbit coupling leads to suppression of
the superconductivity through a continuous phase transition.
Finally we have shown that the Hall conductance tracks the quantum
phase transition induced by magnetic impurities in conventional
superconductors. This provides transport measurement as a possible
tool to detect the transition, related to earlier predictions that
transport properties are affected by the presence of magnetic
impurities in a superconductor  \cite{current}.
We note that one of the interests of the AHE is that it can be easily
measured.

We acknowledge support from FCT grant PTDC/FIS/70843/2006. The
work was partly supported by the Polish Ministry of Science and
Higher Education as research projects in years 2011-2014 (VKD and
JB). This work is also
supported by the cooperation agreement between Poland and Portugal
in the years 2009-2010.


\begin{thebibliography}{99}

\bibitem{hall1880}
E. H. Hall, Philos. Mag. {\bf 10}, 301 (1880); ibid. {\bf 12}, 157 (1881).

\bibitem{chien79}
{\em The Hall Effect and Its Applications}, edited by C. L. Chien and
C. R. Westgate (Plenum Press, New York, 1979).

\bibitem{crepieux01_1}
A. Cr\'epieux and P. Bruno, Phys. Rev. B {\bf 64}, 014416 (2001).

\bibitem{berger72}
L. Berger, Phys. Rev. B {\bf 2}, 4559 (1970); {\bf 5}, 1862 (1972).

\bibitem{smit58}
J. Smit, Physica (Amsterdam) {\bf 24}, 39 (1958).

\bibitem{taguchi01}
Y.~Taguchi and Y.~Tokura, Europhys. Lett. {\bf 54}, 401 (2001); Y.
Taguchi, Y. Oohara, H. Yoshizawa, N. Nagaosa, and Y. Tokura,
Science {\bf 291}, 2573 (2001); S. Onoda and N. Nagaosa, Phys.
Rev. Lett. {\bf 90}, 196602 (2003); Y. Taguchi, T. Sasaki, S.
Awaji, Y. Iwasa, T. Tayama, T. Sakakibara, S. Iguchi, T. Ito, Y.
Tokura, Phys. Rev. Lett. {\bf 90}, 257202 (2003).

\bibitem{bruno04}
P. Bruno, V. K. Dugaev, and M. Taillefumier, \prl {\bf 93}, 096806 (2004).

\bibitem{karplus54}
R. Karplus and J. M. Luttinger, \prl {\bf 95}, 1154 (1954); J. M.
Lutinger, Phys. Rev. {\bf 112}, 739 (1958).

\bibitem{jungwirth02}
T. Jungwirth, Q. Niu, and A. H. MacDonald, \prb {\bf 88}, 207208
(2002); D. Culcer, A. H. MacDonald, and Q. Niu \prb {\bf 68},
045327 (2003);
%T. Jungwirth, J. Sinova, K. Y. Wang, K. W. Edmonds, R. P. Campion,
%B. L. Gallagher, C. T. Foxon, Q. Niu, and A. H. MacDonald,
%Appl. Phys. Lett. {\bf 83}, 320 (2003);
D. Culcer, J. Sinova, N. A. Sinitsyn, T. Jungwirth, A. H.
MacDonald, and Q. Niu, \prl {\bf 93}, 046602 (2004).

\bibitem{m_onoda02}
M. Onoda and N. Nagaosa, J. Phys. Soc. Jpn. {\bf 71} 19 (2002);
\prl {\bf 90}, 206601 (2003); Z. Fang, N. Nagaosa, K. S.
Takahashi, A. Asamitsu, R. Mathieu, T. Ogasawara, H. Yamada, M.
Kawasaki, Y. Tokura, and K. Terakura, Science {\bf 302}, 92
(2003).

\bibitem{nagaosa10}
N. Nagaosa, J. Sinova, S. Onoda, A. H. McDonald, and N. P. Ong,
\rmp {\bf 82}, 1539 (2010).

\bibitem{sinova04}
J. Sinova, T. Jungwirth, J. Cerne, Int. J. Mod. Phys. B {\bf 18},
1083 (2004).

\bibitem{haldane04}
F. D. M. Haldane, \prl {\bf 93}, 206602 (2004).

\bibitem{streda82}
P. Streda, J. Phys. C {\bf 15}, L717 (1982).

\bibitem{rashba84}
Yu. A. Bychkov and E. I. Rashba,
%Pis'ma v Zh. Eksp. Teor. Fiz. {\bf 39}, 64 (1984) [JETP Lett. {\bf 39}, 78 (1984);
J. Phys. C {\bf 17}, 6093 (1984).

\bibitem{nunner07}
T. S. Nunner, N. A. Sinitsyn, M. F. Borunda, V. K. Dugaev, A. A.
Kovalev, A. Abanov, C. Timm, T. Jungwirth, J. Inoue, A. H.
MacDonald, and J. Sinova, \prb {\bf 76}, 235312 (2007).

\bibitem{thouless82}
D. J. Thouless, M. Kohmoto, M. P. Nightingale, and M. den Nijs,
\prl {\bf 49}, 405 (1982).

\bibitem{niu85}
Q. Niu, D. J. Thouless, and Y. S. Wu, \prb {\bf 31}, 3372 (1985).

\bibitem{takahashi02}
S. Takahashi and S. Maekawa, \prl {\bf 88}, 116601 (2002).

\bibitem{Hall} P.D. Sacramento, J. Phys. Cond. Matt. {\bf 11}, 4861 (1999).
\bibitem{Fertig} W.A. Fertig, D.C. Johnston, L.E. DeLong, R.W. McCallum,
M.B. Maple and B.T. Matthias, Phys. Rev. Lett. {\bf 38}, 987
(1977).

\bibitem{Ishikawa} M. Ishikawa and O. Fischer, Solid State Commun. {\bf 23}, 37 (1977).

\bibitem{Felner} I. Felner, U. Asaf, Y. Levi and O. Millo, Phys. Rev. B {\bf 55},
R3374 (1997).

\bibitem{Bernhard} C. Bernhard, J.L. Tallon, Ch. Niedermayer, Th. Blasius,
A. Golnik, E. Brucher, R.K. Kremer, D.R. Noakes, C.E. Stronach and
E.J. Ansaldo, Phys. Rev. B {\bf 59}, 14099 (1999).

\bibitem{Saxena} S.S. Saxena, P. Agarwal, K. Ahilan et. al, Nature {\bf 406},
587 (2000).

\bibitem{Aoki} D. Aoki, A. Huxley, E. Ressouche  et al., Nature {\bf 413}, 613 (2000).

\bibitem{Huy} N.T. Huy, A. Gasparini, D.E. de Nijs, Y. Huang, J.C.P. Klaasse,
T. Gortenmulder, A. de Visser, A. Hamann, T. G\"orlach and H.v.
L\"ohneysen, Phys. Rev. Lett. {\bf 99}, 067006 (2007).

\bibitem{Guanghan} G. Cao, S. Xu, Z. Ren, S. Jiang, C. Feng and Z. Xu, arXiv:1105.3255.

\bibitem{Bauer} E. Bauer, G. Hilscher, H. Michar, Ch. Paul, E. Scheidt,
A. Gribanov, Yu. Seropegin, H. Noel, M. Sigrist and P. Rogl, Phys.
Rev. Lett. {\bf 92}, 027003 (2004).

\bibitem{Samokhine} K. Samokhin, E. Zijlstra and K. Bose, Phys. Rev. B
{\bf 69}, 094514 (2004).

\bibitem{Akazawa} T. Akazawa, H. Hidaka, H. Kotegawa, T. Kobayashi, S. Fukushima,
E. Yamamoto, Y. Haga, H. Settai and Y. Onuki, Physica B {\bf
378-380}, 355 (2006).

\bibitem{Kimura} N. Kimura, Y. Muro and H. Aoki, J. Phys. Soc. Jpn. {\bf 76},
051010 (2007).

\bibitem{Edelstein1} V.M. Edelstein, Phys. Rev. Lett. {\bf 75}, 2004 (1995).

\bibitem{Yip1} S.K. Yip, Phys. Rev. B {\bf 65}, 144508 (2002).

\bibitem{Edelstein2} V.M. Edelstein, Phys. Rev. B {\bf 67}, 020505(R) (2003).

\bibitem{Samokhin} K.V. Samokhin, Phys. Rev. B {\bf 70}, 104521 (2004).

\bibitem{Fujimoto1} S. Fujimoto, Phys. Rev. B {\bf 72}, 024515 (2005).

\bibitem{Hasan} M. Z. Hasan and C. L. Kane
Rev. Mod. Phys. {\bf 82}, 3045 (2010).

\bibitem{Qi} X.-L. Qi and S.-C. Zhang, arXiv:1008.2026.

\bibitem{Fu} L. Fu and C.L. Kane, Phys. Rev. Lett. {\bf 100}, 096407 (2008).

\bibitem{Sau} J.D. Sau, R.M. Lutchyn, S. Tewari and S. Das Sarma, Phys. Rev. Lett.
{\bf 104}, 040502 (2010).

\bibitem{Sato} M. Sato and S. Fujimoto, Phys. Rev. B {\bf 79}, 094504 (2009).

\bibitem{takahashi} S. Takahashi, S. Hikino, M. Mori, J. Martinek, and S.
Maekawa, Phys. Rev. Lett. 99, 057003 (2007).

\bibitem{yokoyama} T. Yokoyama, preprint arXiv:1107.0119, 2011

\bibitem{Sakurai} A. Sakurai, Prog. Theor. Phys. {\bf 44}, 1471 (1970);
A.V. Balatsky, I. Vekhter and J.-X. Zhu, Rev. Mod. Phys. {\bf 78}, 373 (2006).

\bibitem{Morr} M.I. Salkola, A.V. Balatsky and J.R. Schrieffer, Phys. Rev. B
{\bf 55}, 12648 (1997); D.K. Morr and N.A. Stavropoulos, Phys. Rev. B {\bf 67},
020502(R) (2003).

\bibitem{us} P.D. Sacramento, V.K. Dugaev and V.R. Vieira, Phys. Rev. B {\bf 76},
014512 (2007).


\bibitem{QI} P.D. Sacramento, P. Nogueira, V.R. Vieira and V.K. Dugaev,
Phys. Rev. B {\bf 76}, 184517 (2007); N. Paunkovi\'c, P.D. Sacramento, P. Nogueira, V.R. Vieira and V.K. Dugaev,
Phys. Rev. A {\bf 77}, 052302 (2008).

\bibitem{Rashba} L.P. Gor'kov and E.I. Rashba, Phys. Rev. Lett. {\bf 87}, 037004 (2001)

\bibitem{deGennes}
P.G. de Gennes, {\it Superconductivity of Metals and Alloys}
(Addison-Wesley, Reading, MA, 1989).

\bibitem{Sigrist} M. Sigrist and K. Ueda, Rev. Mod. Phys. {\bf 63}, 239 (1991).

\bibitem{Durst} A.C. Durst and P.A. Lee, Phys. Rev. B {\bf 62}, 1270 (2000).

\bibitem{BTK} G. E. Blonder, M. Tinkham, and T. M. Klapwijk,
Phys. Rev B {\bf 25}, 4515 (1982).

\bibitem{Vafek} O. Vafek, A. Melikyan and Z. Tesanovic, Phys. Rev. B
{\bf 64}, 224508 (2001).

\bibitem{Mahan} G.D. Mahan, {\it Many-Particle Physics} (Plenum Press, 1990).

\bibitem{Volovik} G.E. Volovik and L.P. Gor'kov, Sov. Phys. JETP {\bf 61}, 843 (1985).

\bibitem{Frigeri} P.A. Frigeri, D.F. Agterberg, A. Koga and M. Sigrist,
Phys. Rev. Lett. {\bf 92}, 097001 (2004).

\bibitem{Pareek} T.P. Pareek and P. Bruno, Phys. Rev. B {\bf 65}, 241305(R) (2002).

\bibitem{current} P.D. Sacramento, V.K. Dugaev, V.R. Vieira and M.A.N. Ara\'ujo, J.
Phys. Cond. Matt. {\bf 22}, 025701 (2010);
P.D. Sacramento and M.A.N. Ara\'ujo, Eur. Phys. J. B {\bf 76}, 251 (2010);
P.D. Sacramento, L.C. Fernandes Silva, G.S. Nunes,
M.A.N. Ara\'ujo and V.R. Vieira,
Phys. Rev. B {\bf 83}, 054403(2011).





\end{thebibliography}
\end{document}